\documentclass[10pt]{article}
\usepackage{graphicx,amscd,amsmath,amssymb,verbatim}
\usepackage[dvips]{hyperref}
\usepackage{textcomp}
\begin{document}
\title{The alchemy of probability distributions: beyond Gram-Charlier
expansions, and a skew-kurtotic-normal distribution from a rank
transmutation map}
\author{William T. Shaw and Ian R. C. Buckley\footnote{Department of Mathematics, King's College, London; correspondence to william.shaw@kcl.ac.uk}}
\maketitle
\begin{abstract}
Motivated by the need for parametric families of rich and yet
tractable distributions in financial mathematics, both in pricing
and risk management settings, but also considering wider statistical
applications, we investigate a novel technique for introducing
skewness or kurtosis into a symmetric or other distribution.\ We use
a ``transmutation'' map, which is the functional composition of the
cumulative distribution function of one distribution with the
inverse cumulative distribution (quantile) function of another. In
contrast to the Gram-Charlier approach, this is done without
resorting to an asymptotic expansion, and so avoids the pathologies
that are often associated with it. Examples of parametric
distributions that we can generate in this way include the
skew-uniform, skew-exponential, skew-normal, and
skew-kurtotic-normal.
\vskip.2cm
\noindent Presented at the
{\itshape First IMA Computational Finance Conference}. Date: Friday
23rd March 2007.
\end{abstract}

\section{Introduction}

In this paper we consider composite maps of
the following two forms: {\itshape sample transmutation maps} (STMs)
$y=G^{-1}[ F( x) ] $,\ \ %
\noindent and {\itshape rank transmutation maps} (RTMs)
$v=G[ F^{-1}( u) ]$,
where $F$ and $G$ are cumulative distribution functions (CDFs). Such
maps have attracted the attention of statisticians in the past, in
particular of Gilchrist \cite{Gilchrist2000}, who refers to  STMs
and RTMs as $Q$-transformations and $P$-transformations,
respectively.

In this paper our primary focus is the RTM, which we use  as a tool
for the discovery of new families of non-Gaussian distributions. We
use it to modulate a given base distribution for the purposes of
modifying the moments, in particular the skew and kurtosis. An
important example will be to take the base distribution to be
normal, but there is wide latitude in the choice of the base
distribution. An attraction of the approach is that if the CDF and
inverse CDF (or \emph{quantile function} (QF)) are tractable for the
base distribution, they will remain so for the transmuted
distribution.

Applications of the STM include sampling from exotic distributions,
e.g. Student's T \cite{Shaw2006a, ShawLeeCopula, Shaw2006b,
ShawBuckley2007}; and approximating the quantile function for the
normal distribution \cite{Shaw2007a, SteinShaw2007}.

\subsection{Asymptotic approaches}

The Cornish Fisher (CF) series provides a means to approximate the
QF of a distribution in terms of its cumulants. In finance it finds
applications in the calculation of value at risk. Conversely the
Edgeworth or Gram-Charlier (EGC) expansion is used to approximate
the CDF, also based on the cumulants of the distribution. This idea
was first introduced to financial economics by Jarrow and Rudd
\cite{Jarrow1982}, who used the approach to find corrections to the
Black-Scholes price of vanilla options. The CF and EGC series can be
thought of as the asymptotic analogues of the STMs and RTMs,
respectively. For further details see \cite{AbramowitzStegun1972}
(also available on-line at \cite{AbramowitzStegun1972b}) and
\cite{KendallStuartOrd1998}.

Deficiencies and problems with such expansions include: the
requirement that moments be finite, in particular higher moments
when greater accuracy is sought through additional terms in the
expansion, the tendency of the GC method to give negative values for
the PDF, leading to strict bounds on permissible moment values, and
the difficulty in dealing with the special functions that arise.


In common with the asymptotic methods, our goal is to discover
families of parametric distributions with the ability to fit to
levels of third and fourth moments that depart from those of a given
base distribution (e.g. normal, uniform, exponential). However, in
contrast, we do so {\itshape exactly}, and so without encountering
the pitfalls that plague the EGC series.



\subsection{Plan of paper}

Section \ref{XRef-Section-716192147} contains a survey of the skew
normal literature and demonstrations of how to introduce (additional)
skew into the uniform, exponential and normal distributions. In
Section \ref{XRef-Section-71619222} we introduce skew and kurtosis
to the normal distribution. In Section \ref{XRef-Section-716192218}
we conclude.

\subsection{Acknowledgments}

\noindent The authors would like to thank the organisers of the
{\itshape First IMA Computational Finance Conference }and Professors
Samuel Kotz and Marc Genton for helpful comments and explaining
the history and current issues associated with the skew-normal and
related distributions. We also thank Prof Warren Gilchrist for numerous comments on quantile methods.

\section{Rank transmutation, skewness and kurtosis}\label{XRef-Section-716192147}


Given two distributions with a common sample space with CDFs,
$F_{1}$, $F_{2}$, we can define a pair of general RTMs as follows:
\begin{equation}
G_{R_{12}}( u) =F_{2}( F_{1}^{-1}( u) ) , G_{R_{21}}( u) =F_{1}(
F_{2}^{-1}( u) ) .
\end{equation}
The functions $G_{R_{12}}( u) $ and $G_{R_{21}}( u) $ both map the
unit interval $I=[0,1]$ into itself, and under suitable assumptions
are mutual inverses. Naturally they satisfy $G_{R_{i\,j}}( 0) =0$,
and $G_{R_{i\,j}}( 1) =1$. To ensure that transmuted densities are
continuous we optionally make the additional assumption that the
RTMs be continuously differentiable, and in general we require that
the maps be monotone. In the later examples our convention will be
to take $F_{1}$ as the base distribution and $F_{2}$ as the
modulated distribution.

\subsection{Skew-normal literature}


There is an extensive literature on the notion of modulating a given
distribution by the introduction of skewness, dating back to the
work of the early pioneer, Fernando de Helguero (1908)\
\cite{Helguero1908}. A survey of the history of continuous skewed
distributions in general has been made recently by Kotz and Vicari
\cite{Kotz2005}. Another good entry point to the literature is the
article \cite{Azzalini2003}, and an extensive bibliography has been
made available at \cite{AzzaliniRefs}. Azzalini's own recent survey
is available at \cite{Azzalini2005a}, and forms part of a trio of
articles with Genton \cite{Genton2005}, \cite{Azzalini2005b} that
well demonstrates that this is a vigorous area of research. See also
the 2006 paper by Arellano-Valle et al \cite{Arellano-Valle2006}. Briefly, Azzalini {\itshape et al }consider distributions whose
densities are the product of the density $\phi ( x) $ and the
distribution functions $\Phi ( x) $ for some base distribution (e.g.
normal)
\begin{equation}
f( x,\alpha ) =2\phi ( x) \Phi ( \alpha  x) ,
\end{equation}
\noindent where $\alpha $ is a parameter that measures the intensity
of the modulation.
\subsection{Quadratic transmutation}
A natural RTM to consider, which we term the {\itshape quadratic
RTM} (QRTM), has the following simple quadratic form, for $|\lambda
|\leq 1$:
\begin{equation}
G_{R_{12}}( u) =u+\lambda  u( 1-u),
\end{equation}
from which it follows that the CDFs obey the relationship
$F_{2}( x) =\left( 1+\lambda \right) F_{1}( x) -\lambda  {F_{1}( x)
}^{2}$.
Because the inverse RTM is available in closed-form, the sampling
algorithm remains tractable:
\begin{equation}
F_{2}^{-1}( u) =F_{1}^{-1}( G_{R_{21}}( u) ) , \,\,\,
G_{R_{21}}( u)
=\frac{1+\lambda -\sqrt{{\left( \lambda +1\right) }^{2}-4\lambda
u}}{2\lambda }.
\end{equation}
The effect of the QRTM is to introduce skew to a symmetric base
distribution. There is no specific requirement that the base distribution
$F_{1}$ be symmetric. However, if the $F_{1}$ distribution is symmetric
about the origin, in the sense that
$ F_{1}( x) =1-F_{1}( -x)$ ,
we have the result that the distribution of the square
of the transmuted random variable is identical to that of the distribution
of the square of the original random variable. A consequence of
this is that if the original distribution is symmetric, then the
quadratic RTM preserves all even moments. Frameworks within which the distribution of the square is preserved
under the skew transformation are well documented. See for example,
Roberts \& Gesser \cite{Roberts1966}, Gupta \& Cheng \cite{Gupta2004},
Kotz \& Vicari \cite{Kotz2005}, and Wang {\itshape et al }\cite{Wang2004}. We proceed by appling the QRTM to the cases in which the base distribution
is uniform, exponential and normal.
\subsection{Skew-uniform}

We apply the QRTM to the uniform as the base distribution, initially
under the assumption that $|\lambda |\leq 1$
\begin{align}
F_{1}( x) &=x
\\%
F_{2}( x) &=\left\{ \begin{array}{cc}
 0 & x<0 \\
 \left( 1+\lambda \right) x-\lambda  x^{2} & 0\leq x\leq 1 \\
 1 & x>1
\end{array}.\right.
\end{align}
This may be compared with the skew-uniform distribution obtained
within the Azzalini framework that is described in \cite{NadarajahAryal2004}. We can relax the bounds on $\lambda $ if we truncate the modulated
distribution in the following manner:
\begin{equation}
G_{R_{12}}( u) =\min [ \max [ u+\lambda  u( 1-u) ,0] ,1] .
\end{equation}
The PDF for the skew-uniform distribution is shown in Figure
\ref{XRef-Figure-719213022}, top left.
\subsection{Skew-exponential}
To make the point that when adopting the rank transmutation approach
it is not necessary that the base distribution be centred, symmetric
or even defined for negative values, we consider an exponential
base distribution $f_{1}$ for $\beta >0$
\begin{equation}
f_{1}( x) =\left\{ \begin{array}{cc}
 0 & x<0 \\
 \beta   e ^{-\beta  x} & 0\leq x
\end{array},\right.
\end{equation}
which has the transmuted PDF
\begin{equation}
f_{2}( x) =\left\{ \begin{array}{cc}
 0 & x<0 \\
 \beta   e ^{-\beta  x}( 1-\lambda ) +2{\beta \lambda  e }^{-2\beta
x} & 0\leq x
\end{array}.\right.
\end{equation}
With $\beta  = 1$ and $\lambda$ varying from $-1$ to $+1$ in steps
of $1/3$ we obtain the pleasing set of curves shown in Figure
\ref{XRef-Figure-719213022}, top right.
\subsection{Skew-normal }
Of special interest is the case of a normal base distribution,
\begin{equation}
F_{1}( x) =\Phi ( x) :=\frac{1}{2}\left( 1+\operatorname{erf}( \frac{z}{\sqrt{2}})
\right) ,
\end{equation}
where $\phi$ and $\Phi$ are the PDF and CDF for a standard normal
variable, respectively. This gives rise to a transmuted distribution
with density
\begin{equation}
f_{2}( x) =\phi ( x) \left( 1+\lambda -2\lambda  \Phi ( x) \right)
.
\end{equation}
Closed-form expressions for the moments can be found easily and used
for calibration. The standardized PDF (with mean 0 and variance 1) takes the form
\begin{equation}
f_{3}( x) =\sqrt{1-\lambda ^{2}/\pi }f_{2}( x\sqrt{1-\lambda ^{2}/\pi
}-\lambda /\pi ,\lambda ) .
\end{equation}
By expanding in $x$ and $\lambda $ it would be possible to recover
a form of the ECG expansion. However, as our goal is to find similarly
tractable and yet more robust ways to generate exotic distributions
we prefer not to.

\begin{figure}[h]
\begin{center}
\includegraphics[scale=0.8]{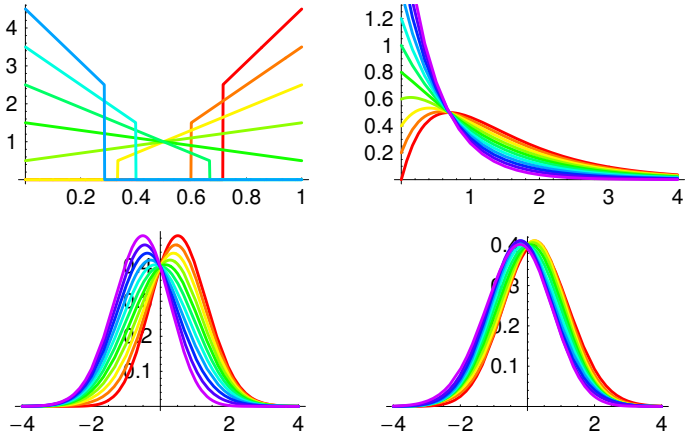}
\end{center}
\caption{PDFs for skew-uniform (top left), skew-exponential (top
right) and skew-normal (bottom) distributions. For the skew-normal
case raw (bottom left) and standardized (bottom right) forms are
shown.}\label{XRef-Figure-719213022}
\end{figure}

The PDFs for the skew-uniform distribution as $\lambda $ is varied
from $-1$ to $+1$ are shown in Figure \ref{XRef-Figure-719213022},
lower two plots. The raw transmuted and normalised versions of the
distributions are shown to the left and the right of the figure,
respectively. By applying the QRTM to other base distributions we can similarly
discover novel skew-Student, skew-Cauchy, and so forth,
distributions. A desirable feature of the approach is that Monte
Carlo sampling via quantile mechanism is as tractable as for the
base distribution in each case.
\subsection{General rank transmutation maps}
Whilst our emphasis so far has been on perturbations of the symmetry
in order to introduce skewness, we can also consider other perturbations.
If we stay within the polynomial structure we can consider maps
of the form
\begin{equation}
G_{R_{12}}( u) =u+u( 1-u) P( u) ,
\end{equation}
\noindent where $P$ is a polynomial with various parameters. It
is of particular interest to place natural constraints on the structure
of $P( u) $.
\subsubsection{Symmetric cubic rank transmutation}
\begin{figure}[h]
\begin{center}
\includegraphics[scale=0.8]{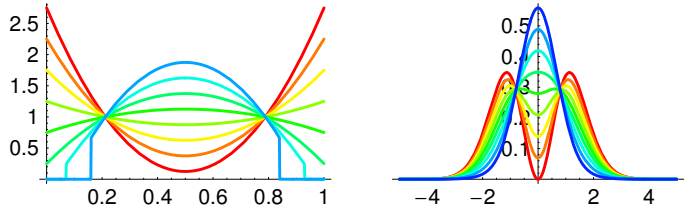}
\end{center}
\caption{Probability density functions obtained when symmetric cubic
rank transformation is applied to uniform (left) and normal (right)
base distributions.}\label{XRef-Figure-719214041}
\end{figure}

The simplest possible type of symmetric mapping is obtained by
choosing $P( u) =\gamma .(u-\frac{1}{2})$ for some constant $\gamma
$. This leads us to define a function that we shall term the
{\itshape symmetric cubic rank transmutation mapping }(SCRTM),
symmetric in the sense that $G_{R_{12}}( 1-u) =1-G_{R_{12}}( u) $.
We could restrict the range of $\gamma $ appropriately but will
project to the unit interval to obtain a map valid for all $\gamma $
as follows:
\begin{equation}
G_{R_{12}}( u) =\min [ \max [ u+\gamma  u( 1-u) \left( u-\frac{1}{2}\right)
,0] ,1] .
\end{equation}
The densities of the distributions that result for various values of
the parameter $\gamma $ are given in Figure
\ref{XRef-Figure-719214041}, for base distributions that are uniform
(left) and normal (right).
\section{Skew-kurtotic transmutations}\label{XRef-Section-71619222}
\subsection{Polynomial family and valid parameter set}
We standardize a family of polynomial rank transmutation maps and
with calibration issues in mind discuss the moment structure. For
parameters $\alpha _{1}$ and $\alpha _{2}$ we consider the polynomial
family
\begin{equation}
P( z,\alpha _{1},\alpha _{2}) =z-z \left( 1-z\right)  \left( \alpha
_{1}+\left( z-\frac{1}{2}\right)  \alpha _{2}\right) ,
\end{equation}
with the restriction on $P$ that it be a monotone
increasing, 1:1 mapping of the unit interval into itself. To achieve
this it is sufficient to impose non-negativity of $P'$ at the
end-points and at $z=1/2$, and that $\min_{0\leq z\leq 1}P'(z)\geq
0$. At particular points in the parameter space $\{\alpha
_{1},\alpha _{2}\}$ the modulated distribution reduces to particular
simple forms. These are listed in Table \ref{XRef-Table-717221246}.
\begin{table}
\caption{Special cases for parameter values $\{\alpha _{1},\alpha
_{2}\}$. Descriptions refer to the relationship between the transmuted
distribution and its base distribution.}\label{XRef-Table-717221246}
\begin{center}
\begin{tabular}{ll}
$\{\alpha _{1},\alpha _{2}\}$ & Description\\
\hline
\{1,0\} & maximum of two\\
\{0,1\} & minimum of two\\
$\{\frac{3}{2},1\}$ & maximum of three\\
$\{-\frac{3}{2},1\}$ & minimum of three\\
\{0,4\} & extreme bimodal\\
$\{0,-2\}$ & middle of three
\end{tabular}
\end{center}
\end{table}

\begin{figure}[hbt]
\begin{center}$
\begin{array}{cc}
\includegraphics[width=1.8in]{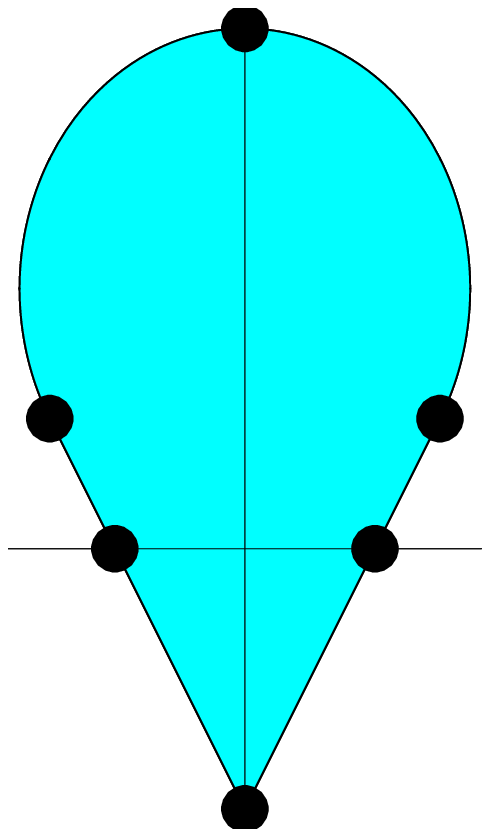} &
\includegraphics[width=2.5in]{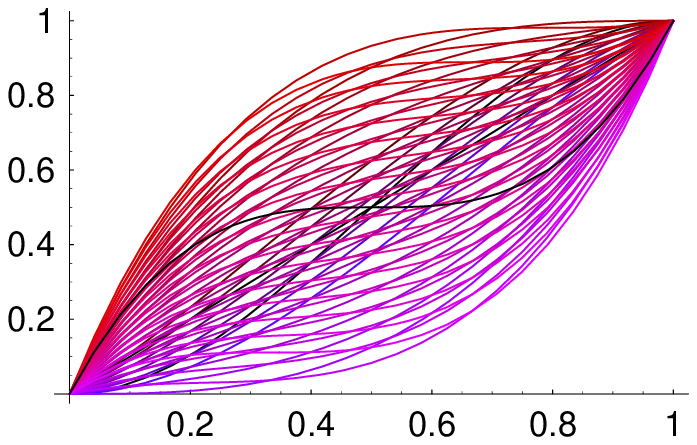}
\end{array}$
\end{center}
\caption{Schematic diagram to show valid parameters in $(\alpha
_{1},\alpha _{2})$-space (left) and permissable skew-kurtotic rank
transmutation maps (right).} \label{Fig:3And4}
\end{figure}

Whilst awkward, these conditions guarantee a globally valid density
function, and the payback is the simplicity of the moment structure.
This region of admissable parameters (i.e. those giving rise to a
well-defined PDF), shown in Figure \ref{Fig:3And4} (left) in
$(\alpha _{1},\alpha _{2})$ space, can be extended by applying a
floor and a cap, but within this region we have a simple polynomial
mapping. Importantly the region contains a large open set around the
origin, which is all that is needed for many practical purposes
where the introduction of a modest amount of skewness and kurtosis
is all that is required. Some typical skew-kurtotic RTMs are shown
in Figure \ref{Fig:3And4} (right).
\subsection{Monte Carlo sampling algorithm}
To simulate using the transmuted distribution it is necessary to
solve the  cubic equation $P( z,\alpha _{1},\alpha _{2}) =u$ for $z$
given $u$. Whilst this can be done in closed form using the approach
of Tartaglia, as described in \cite{Shaw2006}, the logic is
non-trivial. A suitable general cubic solver is described in
\cite{Press2006}. The final step is of course to apply the quantile
function for the base case to the samples of $z$:
$X={F_{1}^{-1}( G_{R_{12}}^{-1}( U) ) }$.
\subsection{Normal case}
When taking $z$ to be the CDF of the normal distribution, the transmuted
density function takes the form
\begin{equation}
F_{2}( x) =\phi ( x)  P'\left( \Phi ( x) ,\alpha _{1},\alpha
_{2}\right) .
\end{equation}
The first few moments when the base distribution is the normal distribution are provided in Table
\ref{XRef-Table-717224333}.
\begin{table}
\caption{First five moments for skew-kurtosis transmutation applied
to a normal base distribution.}\label{XRef-Table-717224333}
\begin{center}
\begin{tabular}{ll}
$k$ & $\mathbb{E}[ X^{k}] $\\
\hline
1 & $\frac{1 }{\sqrt{\pi }}\alpha_1$\\
2 & $1+\frac{\sqrt{3}}{2\pi}\alpha_2$\\
3 & $\frac{5 }{2\sqrt{\pi }}\alpha_1$\\
4 & $3+\frac{13}{2\pi\sqrt{3 }}\alpha_2$
\end{tabular}
\end{center}
\end{table}
\section{Conclusions}\label{XRef-Section-716192218}
Transmutation maps provide a powerful technique for turning the
ranks of one distribution into the ranks of another, e.g. to introduce
skewness in a universal way. These techniques are well adapted for quasi-Monte-Carlo and copula
simulation methods, and may be extended to include a degree of kurtosis,
in contrast to the traditional approach to distributional modulation.
We have given explicit formulae to allow a skew-kurtotic-normal
distribution to be simulated. Proposals for parameter estimation
will be provided in a forthcoming publication. Clearly further work
is needed to
\begin{itemize}
\item make more detailed comparisons with the Azzalini framework;
\item look carefully at the details of the relationship with series
of Gram-Charlier type;
\item identify optimal parameter estimation methods;
\item consider multivariate extensions.
\end{itemize}
However, initial results from our ``alchemy'' studies are very encouraging.
The proposals for skewness adjustments are very simple and may be
applied to any base distribution irrespective of whether it is symmetric
or even defined for $x < 0$. Furthermore, as we have seen, the skewness
adjustments may be extended to manage kurtosis adjustments as well.
Our proposals also contain the basic order statistics (mix, min,
middle) as special cases, and give elegant expressions for the CDFs
of the relevant distributions within a univariate framework. We
are also able to work out moments for the skew-kurtotic-normal developed
within this framework, and these moments are all simple linear functions
of the transmutation parameters. Our techniques are also very well
adapted to Monte Carlo simulation as they make use of the quantile
function of the base distribution composed with an elementary mapping.

\bibliographystyle{amsplain}

\providecommand{\bysame}{\leavevmode\hbox to3em{\hrulefill}\thinspace}
\providecommand{\MR}{\relax\ifhmode\unskip\space\fi MR }
\providecommand{\MRhref}[2]{%
  \href{http://www.ams.org/mathscinet-getitem?mr=#1}{#2}
}
\providecommand{\href}[2]{#2}

\end{document}